\documentclass{aa}

\usepackage{graphicx}
\usepackage{txfonts}
\usepackage{siunitx}
\usepackage{caption}
\usepackage{subcaption}
\usepackage{natbib}
\bibpunct{(}{)}{;}{a}{}{,}
\usepackage[colorlinks=true,linkcolor=blue,citecolor=blue,urlcolor=blue]{hyperref}

\begin{document}

   \title{Improving the light curves of gravitationally lensed quasars with Gaia proper motion data\thanks{The light curves presented in this paper are available at the CDS.}}
   \author{C. Sorgenfrei \and R. W. Schmidt \and J. Wambsganss}
   \institute{Astronomisches Rechen-Institut, Zentrum für Astronomie der Universität Heidelberg,  Mönchhofstrasse 12-14, 69120 Heidelberg, Germany. \email{c.sorgenfrei@stud.uni-heidelberg.de}}
   \abstract{}
   {We show how to significantly improve difference image analysis (DIA) of gravitationally lensed quasars over long periods of time using Gaia proper motions.}
   {DIA requires the subtraction of a reference image from the individual images of a monitoring campaign, using stars in the field to align the images. Since the proper motion of the stars can be of the same order as the pixel size during a several-year campaign, we use Gaia DR3 proper motions to enable a correct image alignment. The proper motion corrected star positions can be aligned by the ISIS package. DIA is carried out using the HOTPAnTS package. We apply point spread function (PSF) photometry to obtain light curves and add a proper motion correction of the PSF star to GALFIT.}
   {We apply our method to the light curves of the three gravitationally lensed quasars HE1104-1805, HE2149-2745 and Q2237+0305 in the $R$ and $V$ band, respectively, obtained using \SI{1}{m} telescopes of the Las Cumbres Observatory from 2014 to 2022. We show that the image alignment and the determination of the lensed quasar positions is significantly improved by this method. The light curves of individual quasar images display intrinsic quasar variations and are affected by chromatic microlensing.}{}
   \maketitle

\section{Introduction}
Multiply imaged quasars produced via the strong gravitational lensing effect are used for various applications in astrophysics and cosmology \citep[see e.g. the recent review][and references therein]{Shajib_2022}. Since the first detection of such a system \citep{Walsh_1979}, about $220$ lensed quasars have been discovered to date\footnote{\url{https://research.ast.cam.ac.uk/lensedquasars/}}. Compact objects in the foreground lensing galaxy close to the line of sight to a quasar image can cause additional microlensing signals detectable as uncorrelated brightness variations in the light curves of the different quasar images (see \citet{Chang_1979} and e.g. reviews \citet{Schmidt_2010,Vernardos_2023}). These microlensing signals can be used to test model predictions for the structure of quasars \citep{Wambsganss_1991,Kochanek_2004,Anguita_2008,Eigenbrod_2008,Poindexter_2010,Morgan_2018}, such as the temperature profile of the accretion disk \citep{Shakura_1973}. In this work, we present a method to obtain long-term quasar light curves using point spread function (PSF) photometry and difference imaging analysis (DIA) with a focus on improving the results by using Gaia proper motion data \citep{GAIA_2016,GAIA_2022}. We note that it is also possible to determine proper motions directly from DIA \citep{Skowron_2014}, but here, we use the Gaia values. We apply this method to three lensed quasars and determine their light curves using data taken at the Las Cumbres Observatory (LCO\footnote{\url{https://lco.global/}}, \citet{Brown_2013}) over \SI{9}{years}. In Sect. \ref{sec:targets} we summarise some properties of our targets before elaborating on our method in detail in Sect. \ref{sec:methods}, present the resulting light curves in Sect. \ref{sec:lightcurves} and conclude in Sect. \ref{sec:conclusion}.

\section{Targets}
\label{sec:targets}
Here we state a few properties of the lensed quasars HE1104-1805, HE2149-2745 and Q2237+0305, which we analysed.

\subsection{HE1104-1805}
\label{sec:he1104}
The lensed double quasar HE1104-1805, discovered by \citet{Wisotzki_1993} in the Hamburg/ESO survey (HES), is located at redshift $z_{\text{Q}}=\SI{2.32}{}$, while the lensing galaxy has a redshift of $z_{\text{L}}=\SI{0.729}{}$ \citep{Lidman_2000}. The two images are separated by \SI{3.19}{arcsec} \citep{Lehar_2000} and display a time delay of $\Delta t_{\text{AB}} = \SI{152.2\pm3.0}{days}$, where image B leads \citep{Poindexter_2007}.
\subsection{HE2149-2745}
\label{sec:he2149}
The quasar HE2149-2745, discovered by \citet{Wisotzki_1996} in HES as well, with $z_{\text{Q}}=\SI{2.033}{}$ and $z_{\text{L}}=\SI{0.603}{}$ \citep{Eigenbrod_2007}, also has two images that are separated by \SI{1.70}{arcsec}. The time delay was determined to be $\Delta t_{\text{AB}} = \SI{103\pm12}{days}$ by \citet{Burud_2002}, where image A leads. \citet{Eulaers_2011} reanalysed the same data, report an additional possible delay and conclude that the given delays are not reliable. Using data taken over \SI{15}{years} by the COSMOGRAIL programme, \citet{Millon_2020} find $\Delta t_{\text{AB}} = {-32.4}^{+18.3}_{-9.3}\SI{}{days}$, which is the value we adopt.

\subsection{Q2237+0305}
\label{sec:q2237}
\begin{figure*}
    \centering
    \includegraphics[width=\textwidth]{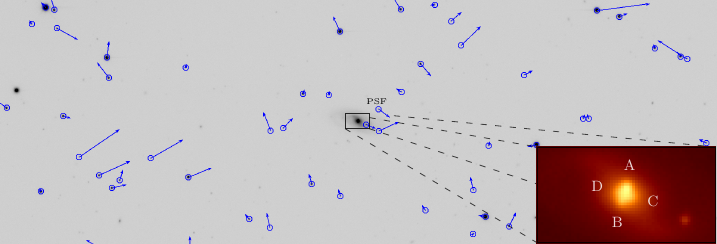}
    \caption{LCO image of the Einstein Cross (Q2237+0305) in the $R$ band in inverted grey scale (north is down, east is right and the field size is $\sim\SI{12.4}{arcmin}\times\SI{4.2}{arcmin}$). Overlayed as blue arrows are the proper motions $\vec{\mu}_i$ of stars in the field from Gaia DR3 data (as reference: the high eastward proper motion star on the top right has $\mu\approx\SI{55.4}{mas/year}$) that we use to improve the image alignment with ISIS (see Sect. \ref{sec:GAIAalign}). The PSF star is marked. In the lower right, a zoom into the central part of the image ($\sim\SI{25.3}{arcsec}\times\SI{13.6}{arcsec}$), where the Einstein Cross is situated, is shown (image names A-D are indicated). Here we use the reference image for difference imaging in the $R$ band.}
    \label{fig:GAIAds9}
\end{figure*}

 Q2237+0305, also known as Einstein Cross, was discovered by \citet{Huchra_1985}. It is visible as four quasar images at $z_{\text{Q}}=\SI{1.695}{}$ close to the center of a barred spiral galaxy known as Huchra's lens (see Fig. \ref{fig:GAIAds9}) at $z_{\text{L}}=\SI{0.0394}{}$. The four images have angular separations of $\lesssim\SI{1.8}{arcsec}$. We use $\Delta t = \SI{0}{days}$ for all time delays between the four images, since estimates (obtained with different techniques) for the multiple time delays are typically of the order of hours \citep[see e.g.][]{Schneider_1988,Wambsganss_1994,Schmidt_1998,Dai_2003,Vakulik_2006,Koptelova_2006,Berdina_2018} in contrast to our light curves over \SI{9}{years}.
 This is the system in which quasar microlensing was first detected \citep{Irwin_1989,Corrigan_1991}.

\section{Methods}
\label{sec:methods}

We want to measure the light curves of the multiple images of three strongly lensed quasars HE1104-1805, HE2149-2745 and Q2237+0305 in the $R$ and $V$ bands, respectively, using data taken at LCO since 2014. LCO is a global network of robotic telescopes and we use data from the \SI{1}{m} telescopes at four locations (mainly Cerro Tololo, Chile; but also: Sutherland, South Africa; Siding Springs, Australia; McDonald, USA). To extract the light curves from the data obtained, the four main steps (which we discuss in more detail in the following four subsections) are:
\begin{enumerate}
    \item[\ref{sec:GAIAalign}] align all images from LCO with respect to each other using the ISIS package\footnote{\url{http://www2.iap.fr/users/alard/package.html}} from \citet{Alard_2000},
    \item[\ref{sec:combining}] combine all images from the same night and produce a high S/N and small seeing reference image for the final step,
    \item[\ref{sec:psfphot}] extract the quasar image A positions and the images PSFs using a modified version of the GALFIT software version 2.0.3 from \citet{Peng_2002} and
    \item[\ref{sec:diffimaging}] apply difference image analysis (i.e. essentially subtracting the reference image transformed with an appropriate convolution kernel from all combined images) using the HOTPAnTS software\footnote{\url{https://github.com/acbecker/hotpants}} by \citet{Becker_2015} which implements the algorithm from \citet{Alard_1998} and \citet{Alard_2000}, and finally extract the light curves by again using GALFIT and therefore applying PSF photometry on the difference images.
\end{enumerate}
Overall, the method is similar to \citet{Giannini_2016}, but we now use Gaia data to improve the astrometry.
We adapted the ISIS, GALFIT and HOTPAnTS codes to deal with specialities of our data such as low number of stars, the multiple images of our quasars, as well as to include improvements using Gaia proper motion data. All steps of our reduction are implemented in python or are python codes that run ISIS, GALFIT and HOTPAnTS on our data. The modified codes (except for GALFIT) and the python codes are available on GitHub\footnote{\url{https://github.com/sorgenfrei-c95/qsoMLdiffcurves}\label{foo:GITHUB}}.

\subsection{Improving the image alignment with Gaia data}
\label{sec:GAIAalign}
In order to successfully combine the single images of each night and to apply difference imaging on these combined images with respect to a reference image, all the single images have to be aligned. To achieve this, ISIS determines the positions of $N\sim 500$ to $1000$ stars in all images of one quasar in one band and with that aligns the images relative to each other by applying small shifts, rotations and scalings to the single images in order to minimize the sum of the squared offsets. But since our images are taken up to $\sim \SI{9}{years}$ apart, the stars positions change due to their proper motion, which makes it difficult for ISIS to achieve proper image alignment.

Here we present our method to improve the alignment significantly. We want to emphasise that the method will be widely applicable and useful in time domain astronomy.

ISIS identifies as many stars as possible in an astrometric reference image and determines all their positions $\vec{r}^{\text{ref}}_i$. It then does the same i.e. determining the star positions $\vec{r}_i$ for all images and minimises all the star deviations to the reference image $\vec{\Delta r}_i=\vec{r}_i-\vec{r}^{\text{ref}}_i$ for each image and corrects that image accordingly. The star positions relative to the reference image are (additionally to image misalignment) influenced by proper motions and the time difference between the current image and the reference image $\Delta t=t-t^\text{ref}$. Since we choose a high signal-to-noise (S/N) and low seeing image of the sample (typically $\sim\SI{1.3}{arcsec}$) as reference image, it is of better quality as most other images, which is why we use it to apply the following correction.

We determine the proper motions from Gaia data release 3 \citep[DR3\footnote{\url{https://gea.esac.esa.int/archive/}}]{GAIA_2022} of stars in our fields with a Renormalised Unit Weight Error $\text{RUWE}<1.4$ \citep{Lindegren_2018} and proper motion values with $\text{S/N}\geq5$ to correct their positions in our reference image in the calculations of ISIS. As preparation, the stars in the neighbourhood of the quasar from the Gaia DR3 data are matched to the stars ISIS detects and intends to use for the alignment (see Fig. \ref{fig:GAIAds9}). This gives us a list of star positions as detected by ISIS, but additionally with their Gaia proper motions in the local tangent plane in right ascension $\mu_{\alpha^*,i}=\mu_{\alpha,i}\cos{\delta_i}$ and declination $\mu_{\delta,i}$. For each image we determine $\Delta t$ and the list of proper motions of all stars and then propagate all star positions in the reference image taken at time $t^\text{ref}$ to the position where they should be at the time of the current image $t$ with
\begin{align}
    \vec{r}^{\text{ref}}_i(t) = \vec{r}^{\text{ref}}_i(t^{\text{ref}}) + \vec{\mu}_i \Delta t \text{~~~with~~~} \vec{\mu}_i =
    (\mu_{\alpha^*,i},\mu_{\delta,i})^\top,
\label{equ:GAIAcorrection}
\end{align}
where the proper motions $\mu_{\alpha^*,i}$ and $\mu_{\delta,i}$ have to be converted to have units of $\SI{}{pixel/years}$ with the pixel scale of $\SI{0.389}{arcsec/pixel}$\footnote{The LCO pixel coordinates $\vec{r}=(x,y)^\top$ used are related to the equatorial coordinates via $(\alpha,\delta)^\top=sR(x,-y)^\top$ with $s$ being the pixel scale and $R$ representing small rotations of less than $\SI{1}{degree}$.}. ISIS then continues unchanged, simply using our corrected star positions $\vec{r}^{\text{ref}}_i(t)$ from Eq. (\ref{equ:GAIAcorrection}) in the reference image to align the current image at time $t$ to the reference image.

\begin{figure*}
    \centering
    \begin{subfigure}[c]{0.5\textwidth}
        \centering
        \includegraphics[width=\textwidth]{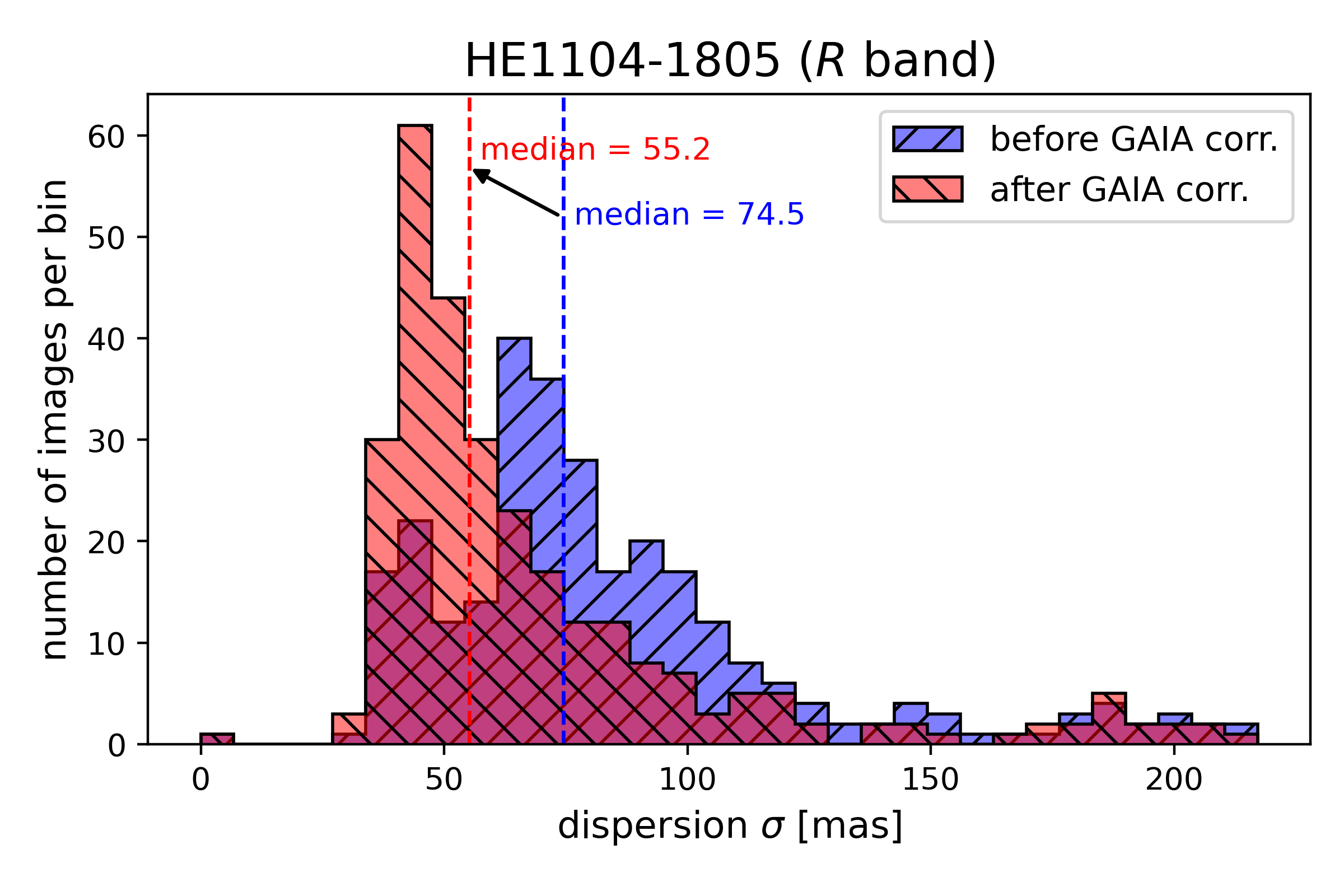}
    \end{subfigure}\hspace*{\fill}
    \begin{subfigure}[c]{0.5\textwidth}
        \centering
        \includegraphics[width=\textwidth]{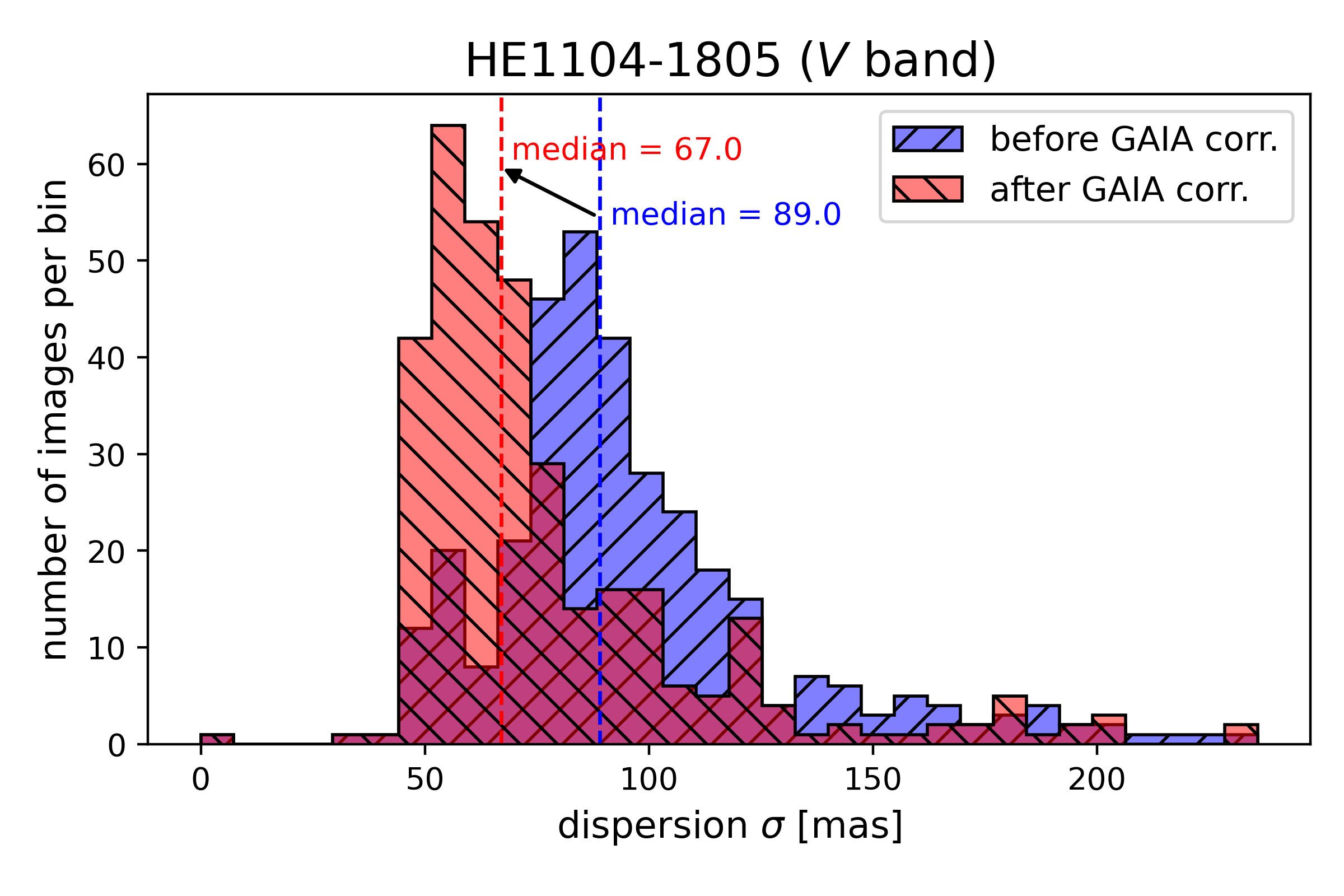}
    \end{subfigure}
    \medskip
    \begin{subfigure}[c]{0.5\textwidth}
        \centering
        \includegraphics[width=\textwidth]{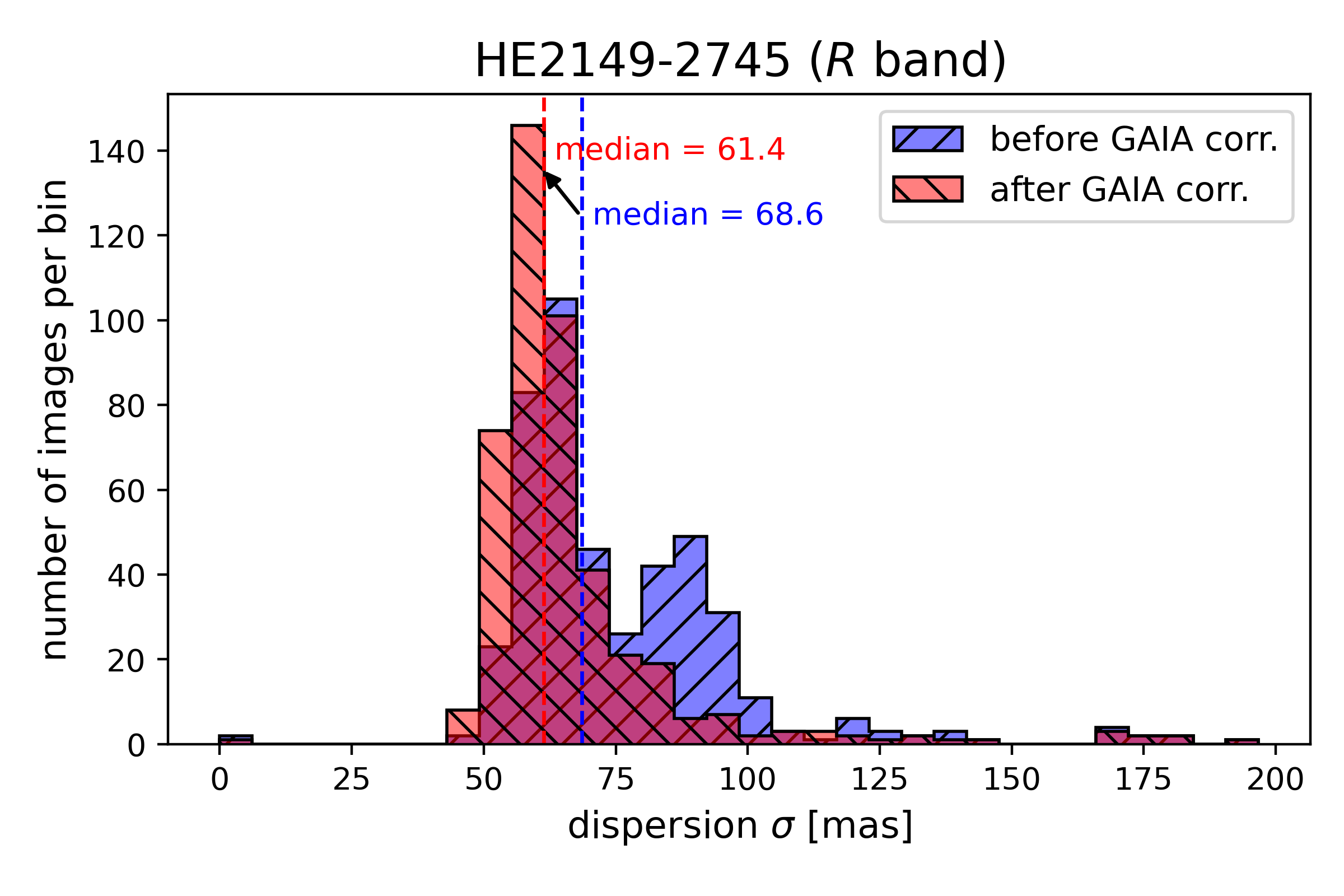}
    \end{subfigure}\hspace*{\fill}
    \begin{subfigure}[c]{0.5\textwidth}
        \centering
        \includegraphics[width=\textwidth]{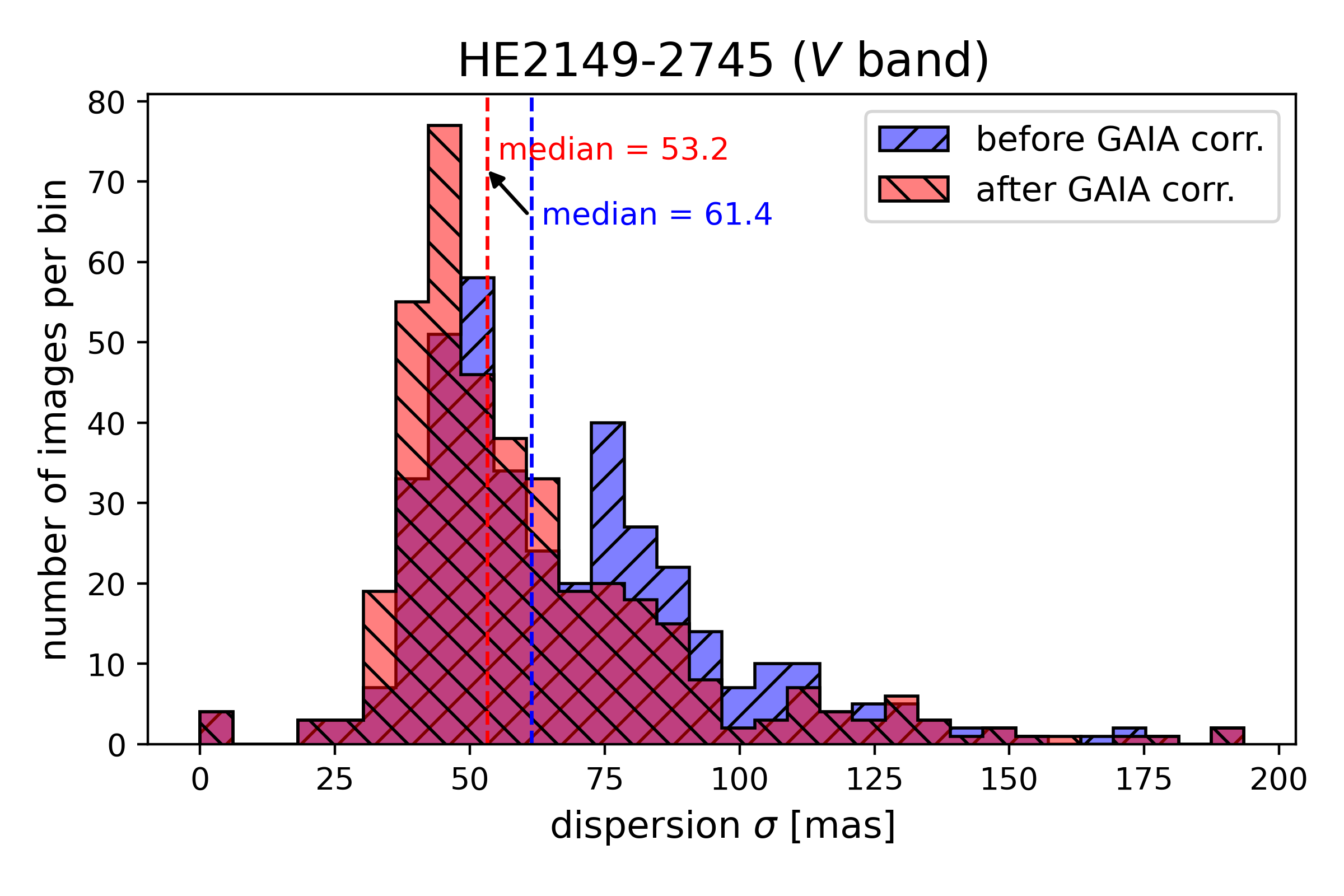}
    \end{subfigure}
    \medskip
    \begin{subfigure}[c]{0.5\textwidth}
        \centering
        \includegraphics[width=\textwidth]{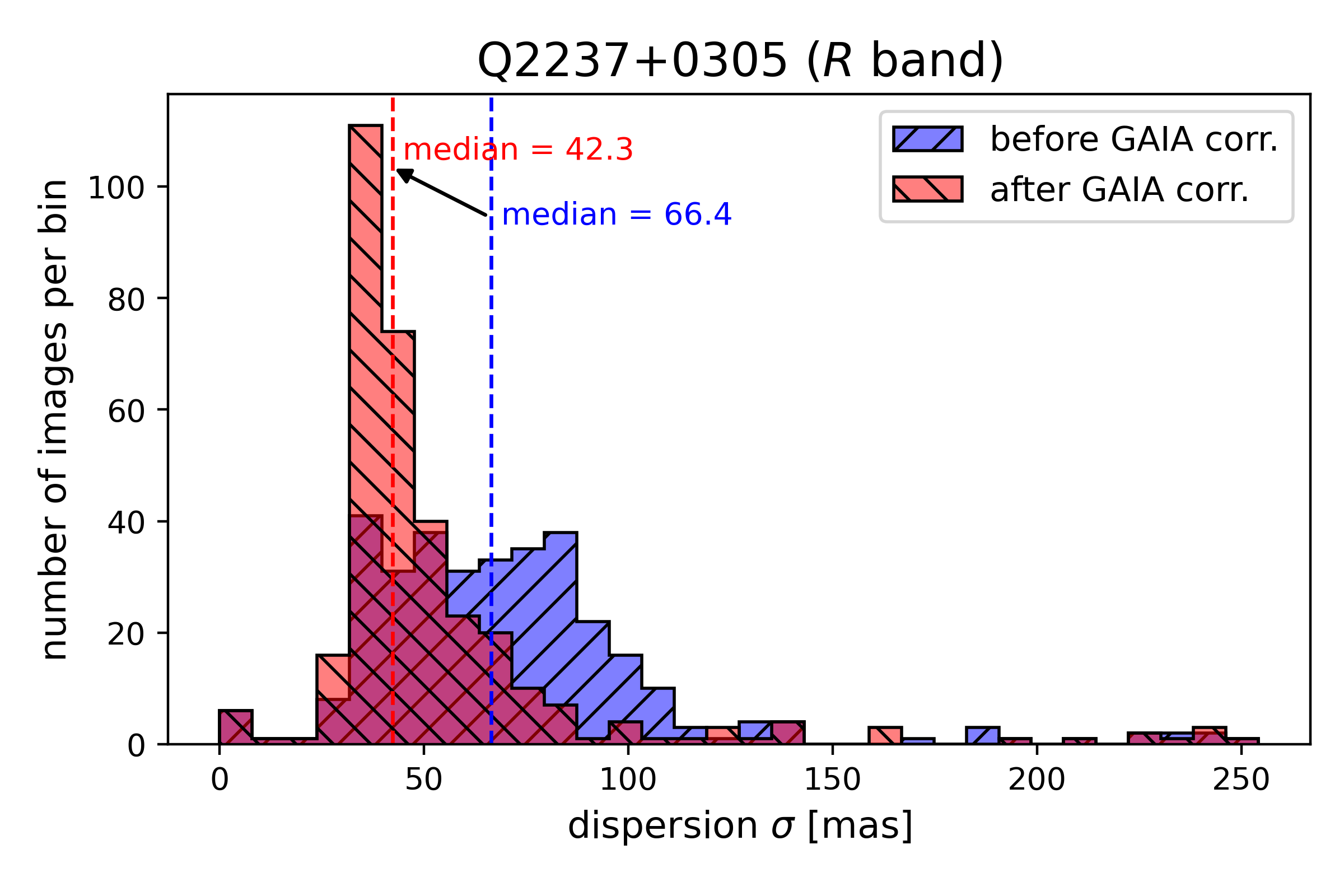}
    \end{subfigure}\hspace*{\fill}
    \begin{subfigure}[c]{0.5\textwidth}
        \centering
        \includegraphics[width=\textwidth]{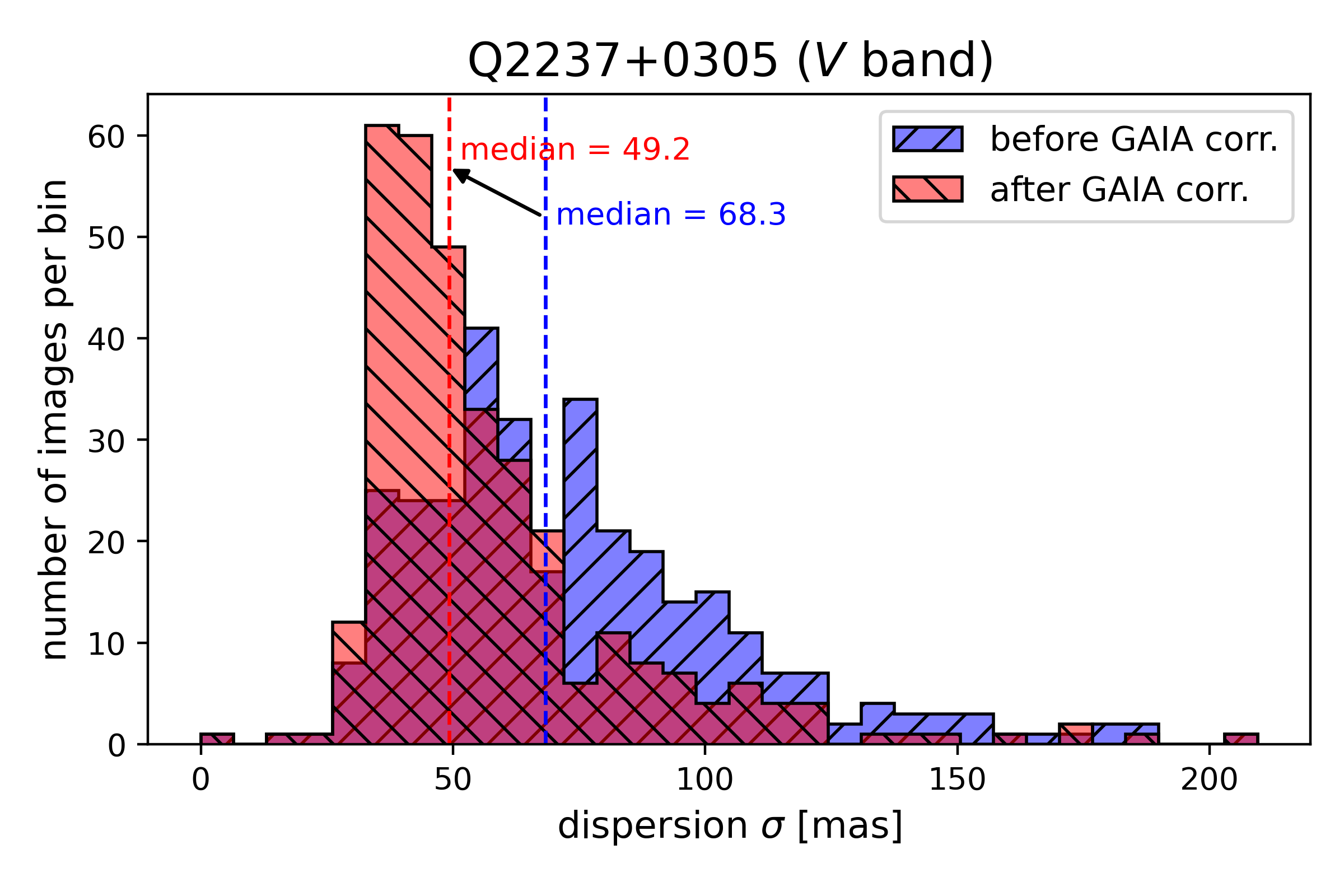}
    \end{subfigure}
    \caption{Results of aligning all the images of HE1104-1805 (first row), HE2149-2745 (second row) and Q2237+0305 (third row) in the $R$ (left column) and $V$ (right column) band. We show the dispersion $\sigma$ (see Eq. (\ref{equ:dispersion})) i.e. the standard deviation of all star position distances in $\SI{}{mas}$ between an aligned image and the reference image for all aligned images without (blue) and with (red) our new method using Gaia proper motion data. The improvement of the median dispersion over all images is shown as well.}
    \label{fig:loginterp}
\end{figure*}

We applied this method to our datasets. To quantify the level of alignment of the images, in Fig. \ref{fig:loginterp} we show the dispersions $\sigma$ of the star position deviations with respect to the reference image $\vec{\Delta r}_i$ after ISIS aligned the images:
\begin{align}
    \sigma = \sqrt{\sigma_x^2+\sigma_y^2} = \sqrt{\frac{1}{N-1}\sum_i^N\left(\vec{\Delta r}_i-\vec{\Delta\bar{r}}\right)^2},
    \label{equ:dispersion}
\end{align}
where $\vec{\Delta\bar{r}}$ is the average star deviation vector that would ideally vanish. The dispersion $\sigma$ for each image is calculated from $\sigma_x$ and $\sigma_y$, the dispersions in $x$ and $y$ pixel-direction as they are determined by ISIS. In Fig. \ref{fig:loginterp} we also compare the results for the dispersions from applying and not applying the correction. The median dispersions are shown as well demonstrating that the level of alignment improved significantly using Gaia.

\subsection{Combining images}
\label{sec:combining}
Typically, three or four images of the same object where taken per night in the same filter with an exposure time of \SI{180}{s} each. These are combined into one image per night (but we do nor combine data from different telescopes) to improve the S/N. In the same step, an error image is calculated, since a high quality error image is needed in order for GALFIT and HOTPAnTS to work properly and reliably in the following steps.

The images are combined, and the associated error image is calculated, using Eqs. (\ref{equ:combining}) and (\ref{equ:combierror}) below, which describe a weighted mean with the possibility of rejecting the $N_{\text{rej}}/2$ minimum and $N_{\text{rej}}/2$ maximum values in every pixel position of the $N$ photometrically scaled single images (min-max-rejection) in order to get rid of bad pixels from e.g. cosmic rays (typically we use $N_{\text{rej}}=2$). The photometric scales $s_i$ measure the relative general brightness and exposure of the images. They are of the order of one and are determined by aperture photometry of multiple ($\sim\SI{20}{}$) reference stars (checked for non-variability over time with respect to each other) in the field via $s_i=\text{median}\left[F^j_1/F^j_i\right]$, where $F^j_i$ is the flux of reference star $j$ in image $i$. The weights $\omega_i$ of the single images are chosen to be the inverse of the photometric scales $s_i$\footnote{The reason for choosing $\omega_i = s_i^{-1}$ is that inverse variance weighting results in the smallest variance of all possible weights of a weighted average. Then, using Poisson statistics, i.e. that the variance in each pixel is given by the flux itself, we find this expression, where we approximate the flux for all pixels by the overall photometric scale of the image, as it is a measure of the typical brightness of the individual images.}. The combined image $\bar{z}$ is given by
\begin{align}
    \bar{z}(x,y)=\frac{\sum_{i=1}^{N-N_{\text{rej}}} \omega_i s_i z_i(x,y)}{\sum_{i=1}^{N-N_{\text{rej}}} \omega_i} \quad\text{with}\quad \omega_i = \frac{1}{s_i},
    \label{equ:combining}
\end{align}
where the $z_i$ are the background $b_i$ reduced single images $d_i$, i.e. $z_i=d_i-b_i$, $i$ counts over the sorted values of $s_i z_i$ where the min-max-rejection is already applied and $(x,y)$ are the individual pixel positions. By using Gaussian error propagation on Eq. (\ref{equ:combining}) and Poisson statistics i.e. $\Delta d_i(x,y)=\sqrt{d_i(x,y)}$, we arrive at the error image given by
\begin{align}
    \Delta\bar{z}(x,y)=\frac{f_{\text{rej}}}{\sum_{i=1}^{N-N_{\text{rej}}} \omega_i}\sqrt{\sum_{i=1}^{N-N_{\text{rej}}} \left[d_i(x,y)+\Delta b_i^2 + \bar{z}(x,y)^2 \frac{\Delta s_i^2}{s_i^4}\right]},
    \label{equ:combierror}
\end{align}
where the correction factor of the min-max-rejection $f_{\text{rej}}$, as well as the background flux error $\Delta b_i$ and the scale error $\Delta s_i$ are determined statistically. For the correction factor, we use $f_{\text{rej}} = \frac{\text{noise of background in combined image}}{\text{median background of error image with}\,f_{\text{rej}}=1}$,
which typically is of the order of one.

\subsection{PSF photometry and extracting quasar image positions improved with Gaia data}
\label{sec:psfphot}
Next, we use PSF photometry with GALFIT on the combined images, not to extract the light curves, but to determine the position of quasar image A and the PSF for every night. They will be necessary when applying PSF photometry with GALFIT to the difference images in the final step (where positions and PSFs cannot be determined reliably). This procedure leads to significantly improved light curves. The PSF is a $\SI{30}{}\times\SI{30}{pixel}$ cutout of a star in the neighbourhood of the quasar we select for GALFIT. We modified GALFIT to include our multiple quasar model, which consists of copies of the PSF arranged in the orientation and fixed relative distances (to quasar image A). The relative coordinates of the images (see Table \ref{tab:HSTdistances}) are taken from the CASTLES webpage\footnote{\url{https://lweb.cfa.harvard.edu/castles/}\label{foo:CASTLES}} (C.S. Kochanek, E.E. Falco, C. Impey, J. Lehar, B. McLeod, H.-W. Rix), which uses Hubble Space Telescope (HST) data \citep{Falco_2001}. GALFIT then fits our quasar model to the combined images by minimizing the sum of squared differences of model flux and data ($\chi^2_{\text{red}}$) returning the fluxes of the images and the quasar position of image A \citep[see also][]{Giannini_2016}.
\begin{center}
\begin{table}
    \sisetup{separate-uncertainty=false}
    \setlength\tabcolsep{2pt}
    \centering
    \caption{Quasar image and galaxy position separations.}
    \begin{tabular}{ c | c c c c }
         & B$-$A & C$-$A & D$-$A & G$-$A \\ \hline
        HE1104-1805, $\Delta\alpha$ & $\hphantom{-}\SI{2.901\pm0.003}{}$ & & &  \\
        HE1104-1805, $\Delta\delta$ & $-\SI{1.332\pm0.003}{}$ & & &  \\
        HE2149-2745, $\Delta\alpha$ & $\hphantom{-}\SI{0.890\pm0.003}{}$ & & &  \\
        HE2149-2745, $\Delta\delta$ & $\hphantom{-}\SI{1.446\pm0.003}{}$ & & &  \\
        Q2237+0305, $\Delta\alpha$ & $-\SI{0.673\pm0.003}{}$ & $\hphantom{-}\SI{0.635\pm0.003}{}$ & $-\SI{0.866\pm0.003}{}$ & $\SI{-0.075\pm0.004}{}$ \\
        Q2237+0305, $\Delta\delta$ & $\hphantom{-}\SI{1.697\pm0.003}{}$ & $\hphantom{-}\SI{1.210\pm0.003}{}$ & $\hphantom{-}\SI{0.528\pm0.003}{}$ & $\hphantom{-}\SI{0.939\pm0.003}{}$
    \end{tabular}
    \label{tab:HSTdistances}
    \setlength\tabcolsep{6pt}
    \sisetup{separate-uncertainty=true}
    \tablefoot{HST measured position separations (in right ascension $\Delta\alpha$ and declination $\Delta\delta$) as obtained from the CASTLES webpage (see \citet{Falco_2001} and footnote \ref{foo:CASTLES}) in $\SI{}{arcsec}$ (with uncertainties of the last digit in parentheses) from quasar image A to image B, C and D, as well as to the lens galaxy G for each quasar, if we used them in the respective GALFIT quasar model.}
\end{table}
\end{center}

\begin{figure}
    \centering\hspace*{\fill}
    \begin{subfigure}[c]{0.155\textwidth}
        \centering
        \includegraphics[width=\textwidth]{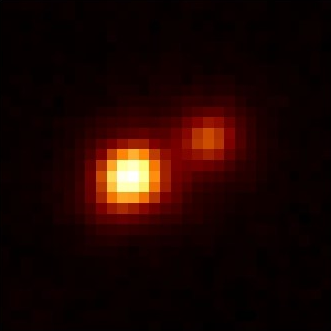}
        \caption{data}
    \end{subfigure}\hspace*{\fill}
    \begin{subfigure}[c]{0.155\textwidth}
        \centering
        \includegraphics[width=\textwidth]{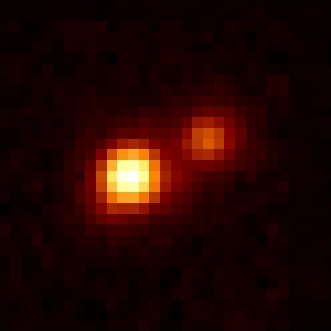}
        \caption{model}
    \end{subfigure}\hspace*{\fill}
    \begin{subfigure}[c]{0.155\textwidth}
        \centering
        \includegraphics[width=\textwidth]{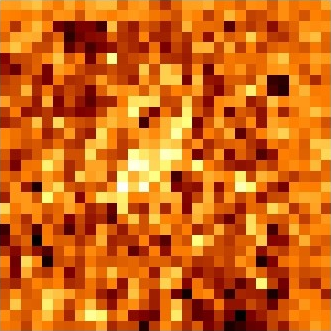}
        \caption{residuals}
    \end{subfigure}\hspace*{\fill}
    \caption{GALFIT example of HE1104-1805 in the $R$ band from July 3, 2019: (a) the combined LCO data of that night, (b) the quasar model made by GALFIT with two copies of the PSF fitted to the two quasar images in the orientation and distance fixed with HST data and (c) the residuals between data and model.}
    \label{fig:he1104GALFIT}
\end{figure}

As an example, in Fig. \ref{fig:he1104GALFIT} we show PSF photometry results of HE1104-1805 with GALFIT. The image separation is $\sim\SI{3.2}{arcsec}$ (see Sect. \ref{sec:he1104}) and the lensing galaxy is not visible in our data (but faintly in e.g. HST images). This works analogously with the quasar HE2149-2745: Again, two images are visible, they have an angular separation of only $\sim\SI{1.7}{arcsec}$ (see Sect. \ref{sec:he2149}) and again the lens galaxy is orders of magnitudes fainter and is not accounted for in our GALFIT model to determine the quasar position and PSF. The third quasar in our dataset, Q2237+0305, i.e. the Einstein Cross, is visible as four images of the quasar on a ring with a radius of $\sim\SI{0.9}{arcsec}$ (see Sect. \ref{sec:q2237}). For this object, the quasar images are barely visible in our data because of the bright lensing galaxy core (see Fig. \ref{fig:GAIAds9}), but well accessible to difference photometry, as we show below.

To adapt the GALFIT model to the Einstein Cross we have to fit four instead of two quasar images and also include the galaxy model described in \citet{Giannini_2017}, which consists of
\begin{enumerate}[(i)]
    \item a de Vaucouleurs profile (i.e. a Sersic profile with fixed sersic index of $n=4$) for the galactic bulge and
    \item an exponential profile ($n=1$) for the galactic disk,
\end{enumerate}
which are both given by the Sersic profile
\begin{align}
I(r)=I_e\exp\left\{-b_n\left[\left(\frac{r}{r_e}\right)^{1/n}-1\right]\right\}\propto
    \begin{cases}
        e^{-b_4\,r^{1/4}} & \text{for } n=4 \\
        e^{-b_1\,r}  & \text{for } n=1,
    \end{cases}
\end{align}
where $I_e=I(r_e)$ and $b_n\approx 2n-1/3$ are chosen such that $r_e$ is the half-light radius. We used the values from \citet{Giannini_2017} to fix the scale radii, the galaxy orientation and the axis ratio, but the relative positions and brightnesses are not fixed. For the average magnitude difference of the galaxy components we find $\sim\SI{0.9}{mag}$ as in \citet{Giannini_2017}.

\begin{figure}
    \centering
    \resizebox{\hsize}{!}{\includegraphics{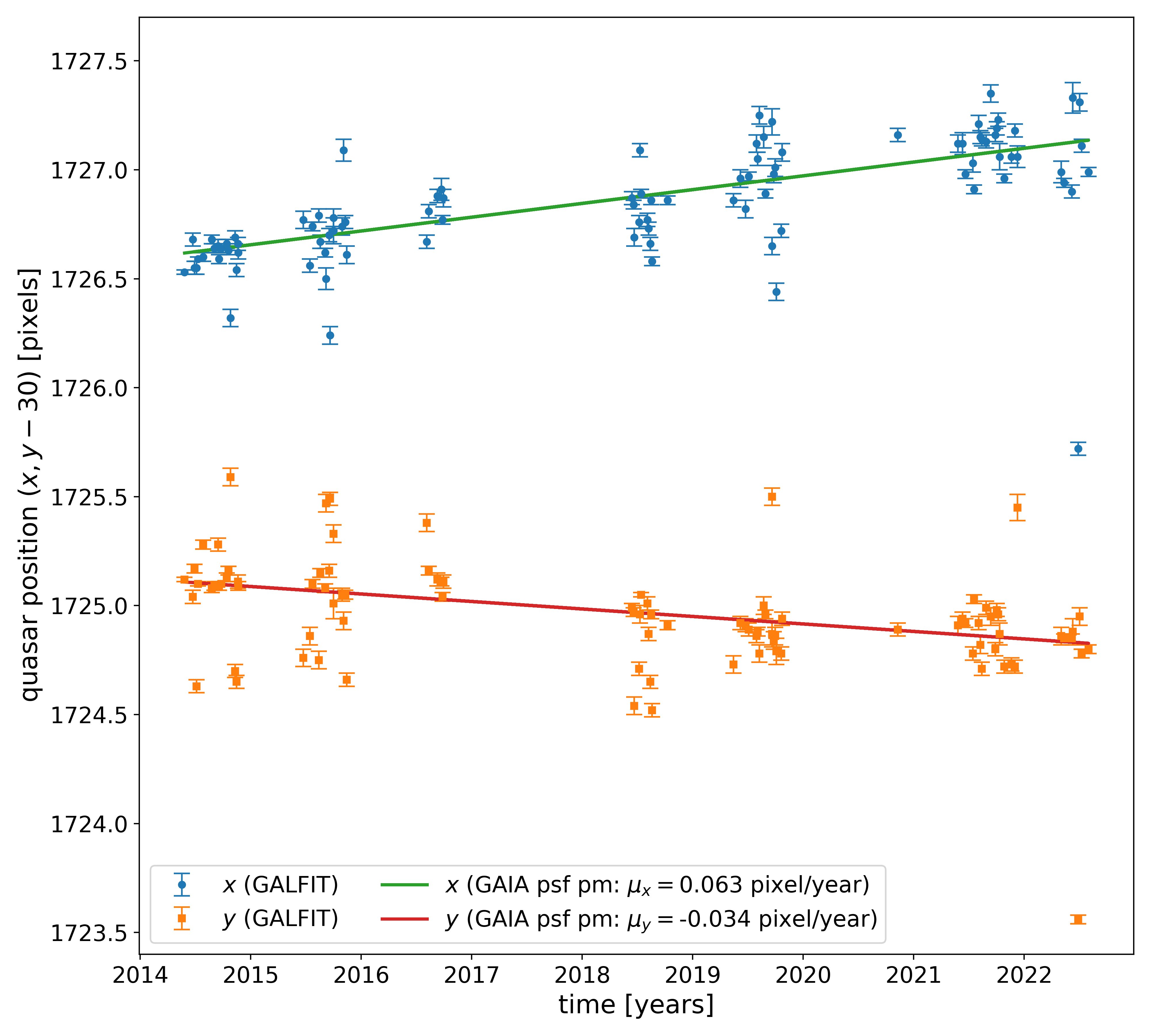}}
    \caption{Q2237+0305 image A $x$ and $y$ pixel positions $\vec{r}_{\text{QSO}}(t)$ as a function of time (in the $R$ band) determined by GALFIT on the combined images shown as upper (blue) and lower (orange) data points respectively. The lines are no fits to these positions, but simply lines $\vec{\bar{r}}_{\text{QSO}}(t)$ with the proper motion $\vec{\mu}_\text{PSF}$ of the PSF star from Gaia as slope and the median of all proper motion corrected positions as overall offset $\vec{\bar{r}}_{0}$ (see Eq. (\ref{equ:betterpospmpsfstar})). We did this for all six datasets and then used the lines as fixed quasar image A positions when applying GALFIT to the difference images to determine the light curves.}
    \label{fig:psfpm_GALFITpos}
\end{figure}

In Fig. \ref{fig:psfpm_GALFITpos} we show the positions of image A of the Einstein Cross as a function of time from 2014 to 2023 as determined with GALFIT in our $R$ data as data points (for the $x$ and $y$ directions respectively). These quasar image A positions $\vec{r}_{\text{QSO}}(t)$ follow lines with non-vanishing slopes. But since we aligned our images (and showed that our method leads to well aligned images), the quasar position should not change with time in our images since the quasar is situated at cosmic distances. The linear trend in the positions comes from the proper motion of the PSF star, which moves inside its fixed 30 by 30 pixel box over the years and therefore the position of this box (fitted to the non-moving quasar) changes with time. The plotted lines are no fits to the data points, but lines with their slopes being calculated from the Gaia DR3 \citep{GAIA_2022} proper motion value for the PSF star $\vec{\mu}_{\text{PSF}}$. The absolute position offset $\vec{\bar{r}}_{0}$ is fixed with the median of all proper motion corrected positions and we use these lines, calculated from Gaia proper motion data of the PSF star, as quasar image A positions:
\begin{align}
    \vec{\bar{r}}_{\text{QSO}}(t) = \vec{\bar{r}}_{0}+\vec{\mu}_{\text{PSF}}\tau(t)~~\text{with}~~\vec{\bar{r}}_{0} = \text{median}\left[\vec{r}_{\text{QSO}}(t)-\vec{\mu}_{\text{PSF}}\tau(t)\right],
    \label{equ:betterpospmpsfstar}
\end{align}
where $\tau(t)=t-\SI{2014.0}{years}$. We use these improved quasar image A positions $\vec{\bar{r}}_{\text{QSO}}(t)$, instead of the data points $\vec{r}_{\text{QSO}}(t)$, since they match the data perfectly. Additionally, this takes the median of the positions, giving a better value, since the (PSF proper motion corrected) quasar position must be constant. This was done for all six datasets (three quasars in the $R$-band and the $V$-band) and the improved quasar positions are kept fixed in the
final determination of the light curves using difference images.

\subsection{Difference imaging}
\label{sec:diffimaging}

\begin{figure*}
    \centering
    \includegraphics[width=\textwidth]{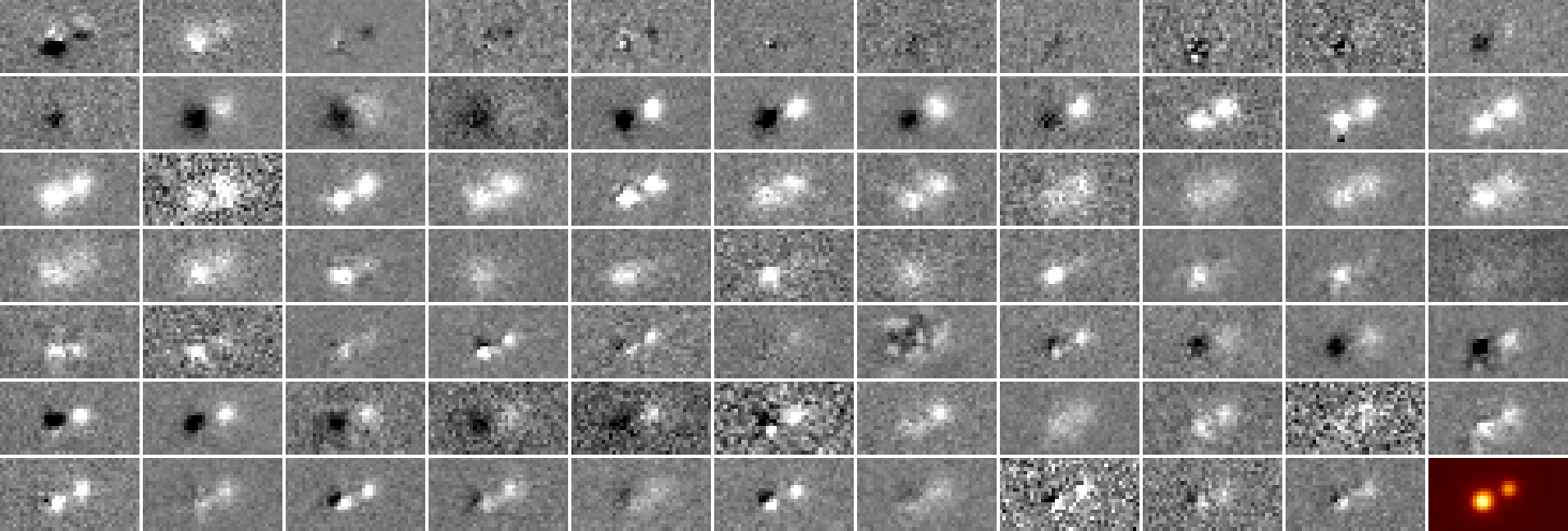}
    \caption{Difference images of HE1104-1805 in the $R$ band from mid 2014 in the upper left to mid 2022 in the lower right (second last image). Shown is the flux in each pixel relative to reference image linearly in grey scale (where white means brighter and black darker), while the last image in the lower right shows the flux in the corresponding reference image logarithmically coloured.}
    \label{fig:diffimages}
\end{figure*}

We also construct a best seeing reference image from $\sim \SI{10}{}$ low seeing images ($\sim\SI{1.3}{arcsec}$) in the same way, as we constructed daily images above (see Appendix \ref{sec:appendixDIA} for more information on the reference images). The difference images are the combined images minus the reference image. However, since the reference image has better seeing than the combined images, HOTPAnTS convolves the reference image with a 
convolution kernel estimated from the PSFs of $\SI{5}{}\times\SI{5}{}$ so-called stamps, which are $\SI{31}{}\times\SI{31}{pixel}$ boxes centered on stars covering the whole image\footnote{The $\SI{5}{}\times\SI{5}{}$ stamp stars are chosen by HOTPAnTS from a list of acceptable stars determined itself by HOTPAnTS, which we can modify to make sure that the quasar images are not included, because this would lead to the quasar disappearing in the difference images.}, as described in \citet{Alard_1998} and \citet{Alard_2000}.
This method requires the specification of the $\sigma$ values for the three Gaussian components of the kernel function. We use \SI{0.8}{pixel}, \SI{2.4}{pixel}, \SI{4.0}{pixel}, respectively, adapted to
the range of seeing values obtained with the LCO telescopes. Moreover, each Gaussian component is multiplied with polynomials of order 4, 3, and 2, respectively. The variation of the
kernel function over the $\SI{5}{}\times\SI{5}{}$ regions of our images is modelled with a spatial polynomial of order 2. Finally, the background is also allowed to vary with a spatial polynomial of
order 2.
The subtraction removes all constant sources in the image and especially removes all the light from the lens galaxy \citep[see also][]{Wozniak_2000a,Wozniak_2000b}. In Fig. \ref{fig:diffimages} we show the difference images and the reference image of HE1104-1805 in the $R$ band.

The remaining variation is due to quasar image brightness variations with respect to the reference image. We measure these with GALFIT (using only the quasar model since the galaxy light has been removed) with the improved quasar positions from the previous step (which are kept fixed), as well as the PSFs. The results are light curves of our quasar images.

One more improvement of the final light curves of the Einstein Cross was achieved by splitting the data into two intervals overlapping in 2018. We only work with reference images of the respective interval and carry out the difference imaging separately. This leads to two light curves for all images that overlap in 2018, which is where we link the brightnesses from times $\geq\SI{2018}{}$ to those $<\SI{2018}{}$. The calibration of the quasar fluxes is described in Sect. \ref{sec:lightcurves}. It can be done consistently with either interval. For the Einstein Cross we used the first interval. The light curves of the Einstein Cross are improved since the stars do not move too far away from their positions in the reference image during these shorter intervals. The difference images from HOTPAnTS then contain less residuals.

\section{Light curves of the lensed quasars HE1104-1805, HE2149-2745 and Q2237+0305}
\label{sec:lightcurves}
Applying the methods described in Sect. \ref{sec:methods}, we reduced our six datasets of LCO data. The number of images in or after the different steps and the final number of epochs used for the light curves are given in Table \ref{tab:imagenumbers}. The main loss of images of about \SI{14}{\%} from all LCO images to the aligned images comes from the visual inspection of the images and subsequent excluding defocused or otherwise damaged images. Typically there are three to four aligned images per night that are then combined into one image as described and for each combined image there is an corresponding difference image (plus one or two reference images for each of our six datasets). In the final light curves, a small number of data points with very large uncertainties were removed (those with $\sigma>\!\SI{0.05}{mag}$ in HE1104-1805 and $\sigma>\!\SI{0.04}{mag}$ in HE2149-2745). The outliers were identified in a logarithmic histogram of the measured uncertainties. 
\begin{center}
\begin{table}
    \centering
    \caption{Number of images after each reduction step.}
    \begin{tabular}{ c | c c c c }
         & LCO & align. & \!comb./diff.\! & \!light curve\! \\ \hline
        HE1104-1805, $R$ & 353 & 273 & 76 & 76\\
        HE1104-1805, $V$ & 370 & 317 & 89 & 86\\
        HE2149-2745, $R$ & 488 & 446 & 126 & 124\\
        HE2149-2745, $V$ & 423 & 385 & 108 & 103\\
        Q2237+0305, $R$ & 384 & 326 & 94 & 94\\
        Q2237+0305, $V$ & 386 & 313 & 93 & 93
    \end{tabular}
    \tablefoot{Number of images remaining after each reduction step, which are original images from LCO, aligned images after sorting out bad images, combined images and difference images from every observation night and finally number of nights used for the light curves, excluding a small number of data points with very large uncertainties (see Sect. \ref{sec:lightcurves}).}
    \label{tab:imagenumbers}
\end{table}
\end{center}

\begin{figure*}
    \centering
    \begin{subfigure}[c]{0.5\textwidth}
        \centering
        \includegraphics[width=\textwidth]{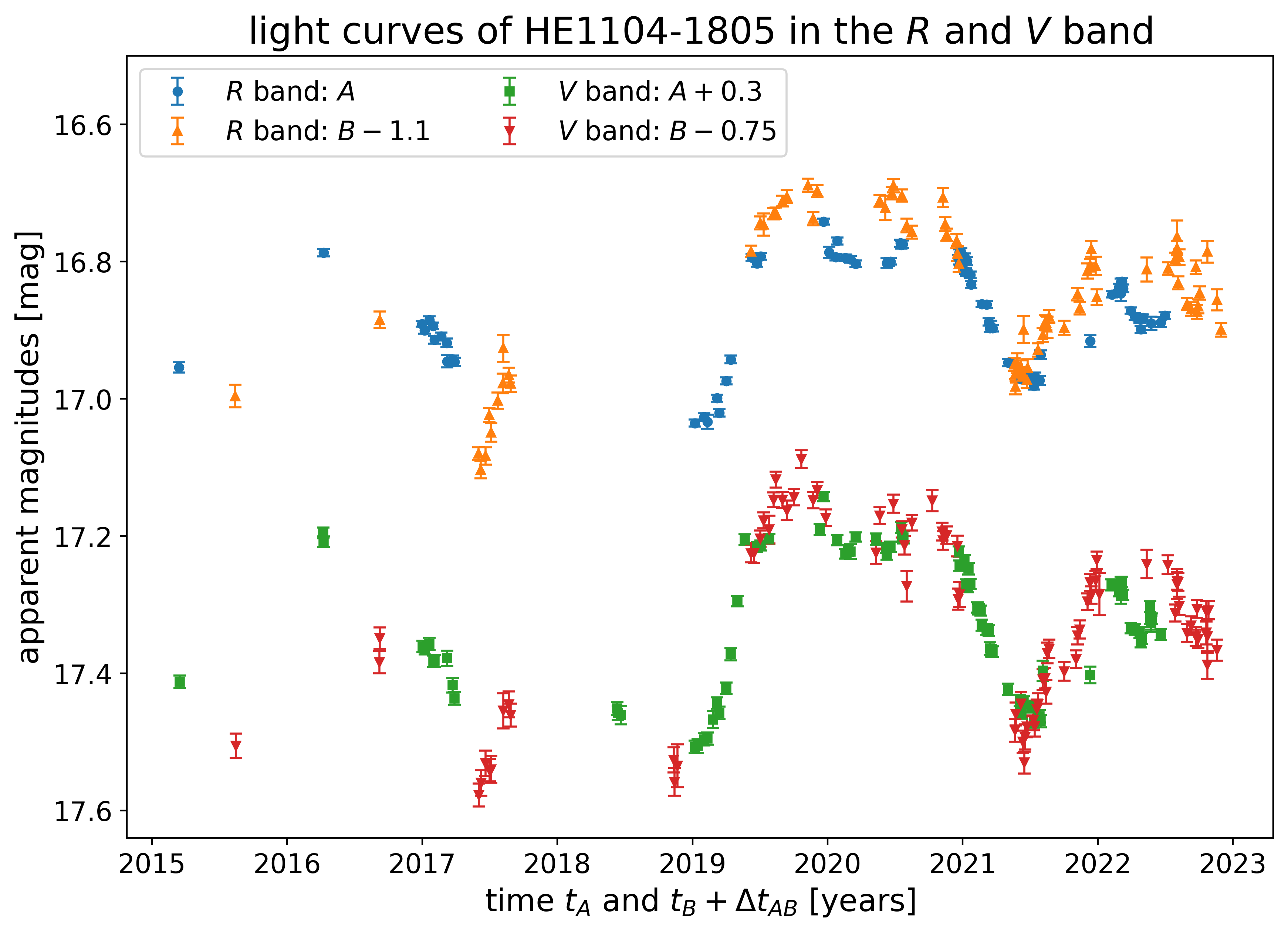}
    \end{subfigure}\hspace*{\fill}
    \begin{subfigure}[c]{0.5\textwidth}
        \centering
        \includegraphics[width=\textwidth]{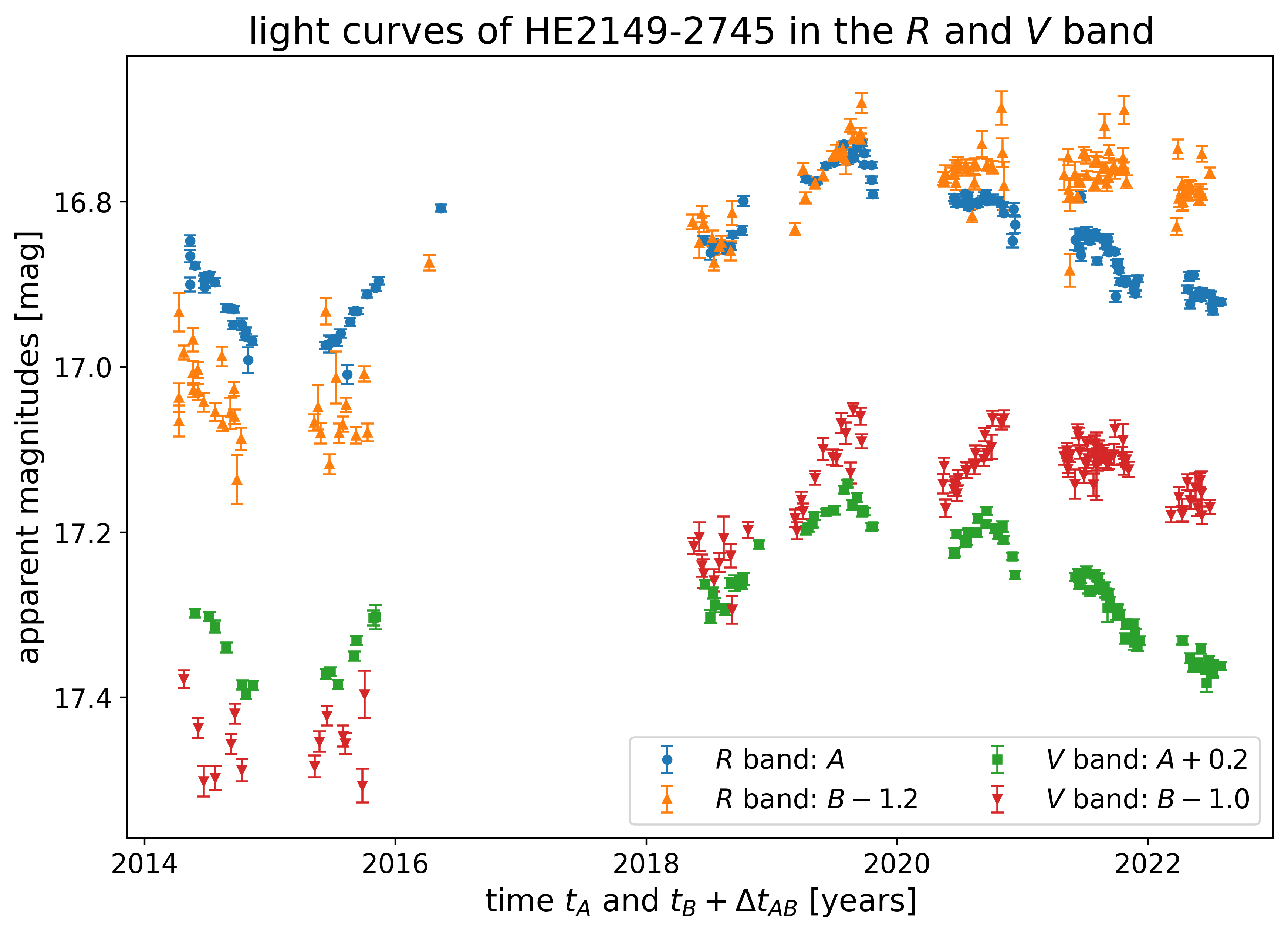}
    \end{subfigure}
    \medskip
    \begin{subfigure}[c]{0.5\textwidth}
        \centering
        \includegraphics[width=\textwidth]{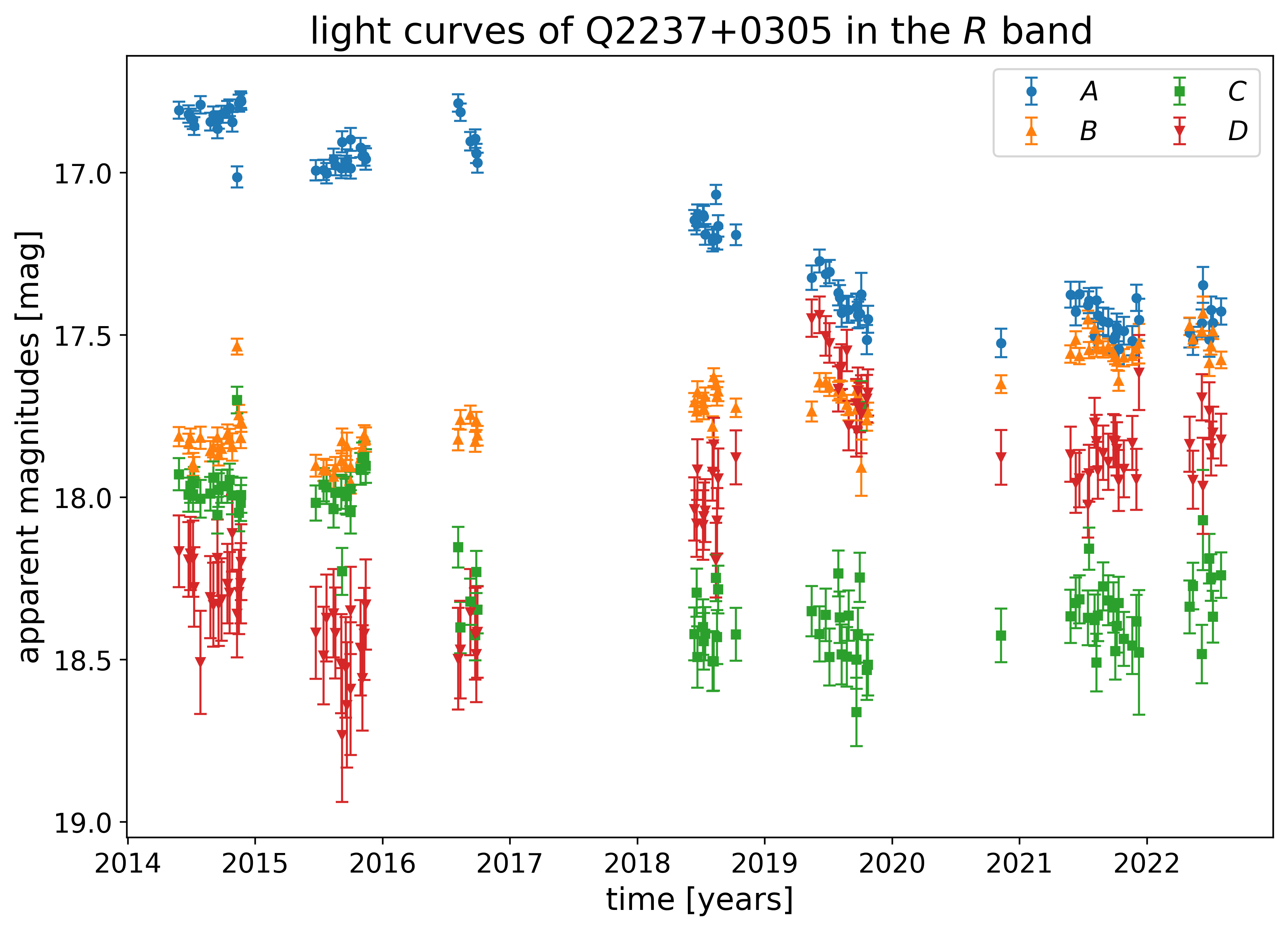}
    \end{subfigure}\hspace*{\fill}
    \begin{subfigure}[c]{0.5\textwidth}
        \centering
        \includegraphics[width=\textwidth]{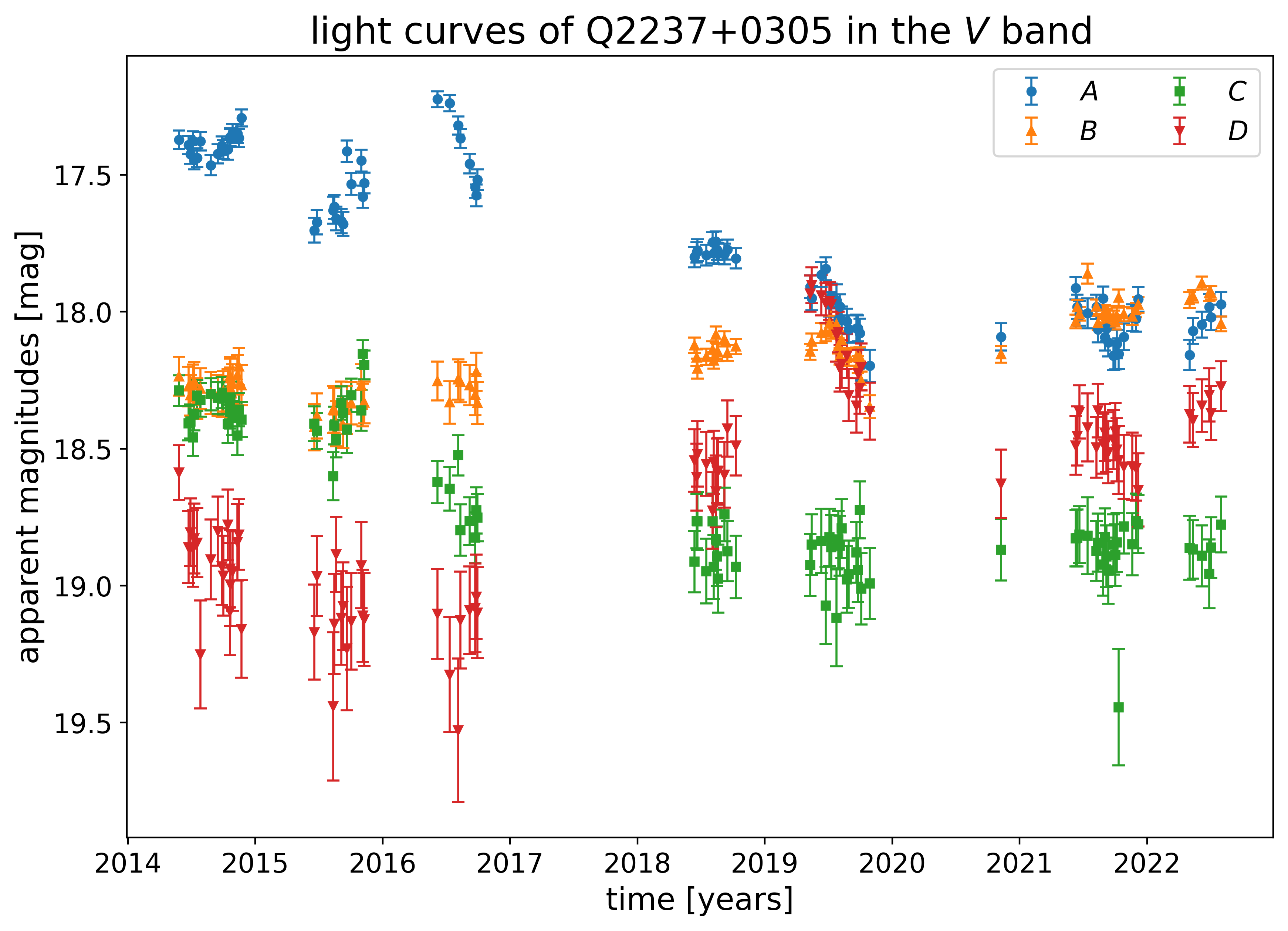}
    \end{subfigure}
    \caption{Final light curves. Upper row: time delay corrected light curves of the images of HE1104-1805 and HE2149-2745 in the $R$ and $V$ bands together respectively, including offsets of the apparent magnitudes for the individual images for displaying reasons. Lower row: light curves of the four images of Q2237+0305 separated into $R$ and $V$ bands with no time delays or magnitude offsets included.}
    \label{fig:lightcurves}
\end{figure*}

Figure \ref{fig:lightcurves} shows the light curves i.e. the apparent magnitudes with time of the multiple quasar images extracted from the difference images with GALFIT as described in Sect. \ref{sec:diffimaging}. The quasar brightness in the reference images (for each quasar and each band) are determined by running GALFIT (using the multiple quasar model and additionally including the galaxy model only for the Einstein cross). The zero points are determined from multiple ($\sim 20$) reference stars with known brightnesses. The apparent magnitudes of the reference stars used to determine the zero points are calculated from the Gaia magnitudes $G$, $G_{bp}$ and $G_{rp}$ using the conversion formulas to the $R$ and $V$ band from Table C.2 in \citet{Riello_2021}.

Finally, the light curves of the images of the double quasars HE1104-1805 and HE2149-2745 are time delay corrected (see Sects. \ref{sec:he1104} and \ref{sec:he2149}), as well as shifted with magnitude offsets for displaying reasons to separate the two bands. They show the intrinsic variations of the quasar as well as deviations between the images indicating microlensing events. We can compare the LCO light curve for HE2149-2745 with the $R$ band light curve from \citet{Millon_2020} because the observing intervals overlap in the years from 2014 to 2016. They are in good agreement, although the fainter image B displays larger scatter in the LCO data.

The light curves of the four images of Q2237+0305 have time delays $\lesssim\SI{1}{day}$ (see Sect. \ref{sec:q2237}) and are presented here with $\Delta t = \SI{0}{days}$ and no relative magnitude offsets. The brightness variations in the images of the Einstein Cross appear mostly different and uncorrelated (except e.g. for a magnitude decline in 2019) indicating microlensing affects all images all the time \citep[see e.g.][and references therein]{Udalski_2006,Millon_2020}.

\begin{figure*}
    \centering
    \includegraphics[width=\textwidth]{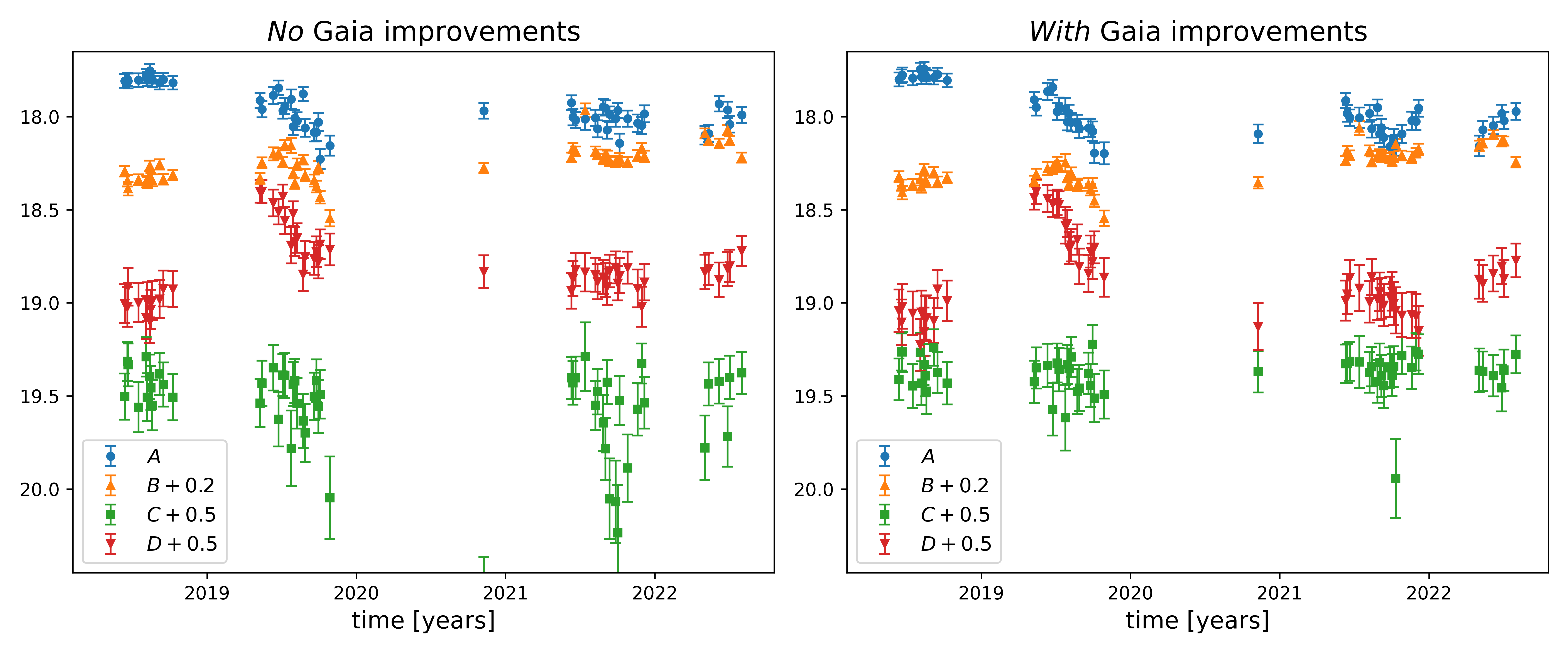}
    \caption{Light curves of Q2237+0305 in the $V$ band without (left panel) and with (right panel) the combined effects of the Gaia proper motion data. We added magnitude offsets to the different images (but the same for both plots) to separate the light curves.}
    \label{fig:compNOvsWITHgaia}
\end{figure*}

We have used Gaia proper motion data to improve the image alignment (see Sect. \ref{sec:GAIAalign} and Fig. \ref{fig:loginterp}) and the quasar position (see Sect. \ref{sec:psfphot} and Fig. \ref{fig:psfpm_GALFITpos}). The effect of these improvements on the quasar light curves is illustrated in Fig. \ref{fig:compNOvsWITHgaia}. The left panel and right panel compare the light curves of Q2237+0305 in the $V$ band without and with these two Gaia steps. We note that we concentrate on the interval from 2018 to 2022 with the relevant second reference image. The effect is stronger for the fainter images (especially image C which is the faintest in that period of time), but also for images A and B the signal was improved.

\section{Conclusion}
\label{sec:conclusion}
Using data taken at LCO from 2014 to 2022, we have demonstrated our method to carry out DIA on long-term monitoring data of the strongly lensed quasars HE1104-1805, HE2149-2745 and Q2237+0305 in the $R$ and $V$ bands.
The method makes use of Gaia DR3 proper motion data of field stars to improve (a) the image alignment as described in Sect. \ref{sec:GAIAalign} and (b) the quasar position (which needs to be fixed for PSF photometry using an observed PSF on the difference images) as described in Sect. \ref{sec:psfphot}.

Additionally, we find that (depending on the object) it can be helpful to split the data into two or more intervals with separate reference images to conduct the difference imaging. The resulting photometry can be combined afterwards where the intervals overlap (in our case: two \SI{5}{year} intervals with \SI{1}{year} overlap for the data of the Einstein Cross). This works only if there are enough high quality images to create reference images for the different intervals. When is this step necessary? One has to consider that our method aligns the images based on the hypothetical star positions at the epoch of the reference image. HOTPAnTS, however, can only work with the observed data, and so the convolution kernel will be affected by this. In the case of Q2237+0305, we find that using two intervals leads to a consistent photometry over the whole observing campaign.

There are many current applications for difference imaging over long periods of time. A prominent example is the upcoming survey of the Vera C. Rubin Observatory. Whenever stars are used as a reference to align images from different nights/epochs the method introduced here to use Gaia proper motion data for improved difference imaging can be useful. Especially whenever PSF photometry is carried out on difference imaging data where the position of the objects needs to be fixed, one has to take care to include the proper motion of the PSF star.

\begin{acknowledgements}
We would like to thank the anonymous referee for a very helpful report.
This work has made use of data from the European Space Agency (ESA) mission {\it Gaia} (\url{https://www.cosmos.esa.int/gaia}), processed by the {\it Gaia} Data Processing and Analysis Consortium (DPAC, \url{https://www.cosmos.esa.int/web/gaia/dpac/consortium}). Funding for the DPAC has been provided by national institutions, in particular the institutions participating in the {\it Gaia} Multilateral Agreement.
We acknowledge the use of Astropy (\url{http://www.astropy.org}).
C.S. acknowledges support from the International Max Planck Research School for Astronomy and Cosmic Physics at the University of Heidelberg.
This work makes use of observations from the Las Cumbres Observatory global telescope network.
We thank Markus Hundertmark and Yiannis Tsapras for helpful discussions and feedback.
\end{acknowledgements}

\bibliographystyle{aa}
\bibliography{BibTeX_CS}

\begin{appendix}
\section{Remarks on the DIA reference images}
\label{sec:appendixDIA}
We used images with low seeing and background from the following dates (in the format YYMMDD) for constructing the reference images for the difference image analysis (Sect. \ref{sec:diffimaging}):
\begin{itemize}
    \item[$\bullet$] HE1104-1805, $R$: 161231, 170120, 170130, 210311, 220410
    \item[$\bullet$] HE1104-1805, $V$: 170131, 210219, 220523
    \item[$\bullet$] HE2149-2745, $R$: 150621, 150902, 210926, 220630
    \item[$\bullet$] HE2149-2745, $V$: 150717, 180903, 200728, 210805,\\\hspace*{74pt}210811, 211005, 220703
    \item[$\bullet$] Q2237+0305 (1), $R$: 140526, 140926, 160926
    \item[$\bullet$] Q2237+0305 (1), $V$: 140526, 141014, 141020, 160907
    \item[$\bullet$] Q2237+0305 (2), $R$: 180713, 210620
    \item[$\bullet$] Q2237+0305 (2), $V$: 180816, 210610
\end{itemize}
Here, (1) and (2) refer to the first and second time intervals into which the data of Q2237+0305 was split for DIA (see Sect. \ref{sec:diffimaging}). We note that in general, not all images of one date were used for the reference images. For detailed lists of all LCO filenames from which the reference images were constructed, also containing seeing and background values, scales and weights (i.e. $s_i$ and $w_i$ from Sect. \ref{sec:combining}), as well as exposure times, see footnote \ref{foo:GITHUB}.
\end{appendix}

\end{document}